\UseRawInputEncoding
\documentclass[prl,twocolumn,superscriptaddress,showpacs,floatfix,reprint]{revtex4-1}

\usepackage{epsfig,amsmath,amssymb,latexsym}

\usepackage{graphicx}
\usepackage{subfigure}
\usepackage[normalem]{ulem}
\usepackage{soul}
\usepackage{color,xcolor,xspace}
\usepackage[colorlinks,
            linkcolor=navy,
            anchorcolor=navy,
            citecolor=navy,
            urlcolor=navy
            ]{hyperref}
\definecolor{red}{rgb}{1,0,0}
\definecolor{blue}{rgb}{0,0,1}
\definecolor{black}{rgb}{0,0,0}
\definecolor{navy}{RGB}{0,0,128}

\begin{document}

\title{Complex structure due to As bonding and interplay with electronic structure in superconducting BaNi$_2$As$_2$}

\author{Bing-Hua Lei}
 \affiliation{Department of Physics and Astronomy, University of Missouri, Columbia, Missouri 65211-7010, USA}

\author{Yucheng Guo}
\affiliation{Department of Physics and Astronomy, Rice University, Houston, Texas 77005, USA}

\author{Yaofeng Xie}
\affiliation{Department of Physics and Astronomy, Rice University, Houston, Texas 77005, USA}

\author{Pengcheng Dai}
\affiliation{Department of Physics and Astronomy, Rice University, Houston, Texas 77005, USA}

\author{Ming Yi}
\affiliation{Department of Physics and Astronomy, Rice University, Houston, Texas 77005, USA}

\author{David J. Singh}
 \email{singhdj@missouri.edu}
 \affiliation{Department of Physics and Astronomy, University of Missouri, Columbia, Missouri 65211-7010, USA}
 \affiliation{Department of Mechanical and Aerospace Engineering, University of Missouri, Columbia, MO 65211, USA}
 \affiliation{Department of Chemistry, University of Missouri, Columbia, MO 65211, USA}

\date{\today}

\begin{abstract}
BaNi$_2$As$_2$ is a superconductor chemically related to the Fe-based superconductors, with a complex and
poorly understood structural phase transition.
We show based on first principles calculations that in fact there are two distinct competing structures. 
These structures are very different from electronic, transport and bonding points of view
but are close in energy.
These arise due to complex As bonding patterns and drive distortions of the Ni layers.
This is supported by photoemission experiments.
This leads to an interplay of electronic and structural behavior including induced anisotropic of the
electronic transport.
The competition between these distortions is associated with
the complex behavior observed in BaNi$_2$As$_2$ samples.
\end{abstract}

\maketitle

The discovery of iron-based superconductivity
\cite{Kamihara_2008}
led to
extensive activity exploring the behavior of these and related compounds.
This resulted in the discovery of remarkably and unexpectedly rich behavior,
including apparently distinct types of superconductivity,
various manifestations of correlated electron behavior,
quantum critical phenomena
and strong interplays between correlated electrons and magnetism,
and
different symmetry breakings,
including electronic nematic behavior and structural transitions
\cite{johnston,Mazin_2008,dai,si,worasaran}.
Unraveling these interplays may be key to general understanding of high temperature
superconductivity and correlated electron behavior.

Superconducting BaNi$_2$As$_2$ is a particularly interesting case
\cite{Ronning_2008}.
BaNi$_2$As$_2$ occurs in the ideal tetragonal ThCr$_2$Si$_2$ structure at ambient temperature,
similar to BaFe$_2$As$_2$
\cite{pfisterer}.
However, below $\sim$130 K a lower symmetry structure is found.
This low temperature structure is superconducting, with bulk fully gapped superconductivity based on specific
heat measurements
\cite{kurita}.
Sefat and co-workers, using single crystal x-ray studies classified this structure as monoclinic or triclinic,
with best fit for
triclinic.
The resulting structure had a distortion from tetragonal in the Ni planes
to form zigzag chains of Ni atoms with shorter bond lengths of $\sim$ 2.8 \AA, as compared to the
interchain Ni distances of $\sim$ 3.1 \AA, and the Ni-Ni distance in the ideal tetragonal structure of 2.93 \AA.
This was discussed as ordering associated with Ni orbital fluctuations based on tight binding models
\cite{yamakawa}.
The transition is accompanied by a very strong signature in the resistivity
\cite{Sefat_2009}.
Photoemission measurements showed that the transition is not of magnetic character,
consistent with the susceptibility data \cite{Sefat_2009} and that it is strongly first order, with
considerable hysteresis in band shifts \cite{Zhou_2011}.
This is consistent
with neutron diffraction studies where a large hysteresis of $\sim$10 K was found
\cite{kothapalli}.

\begin{figure*}[htbp]
 \centering
 \includegraphics[width=1.3\columnwidth]{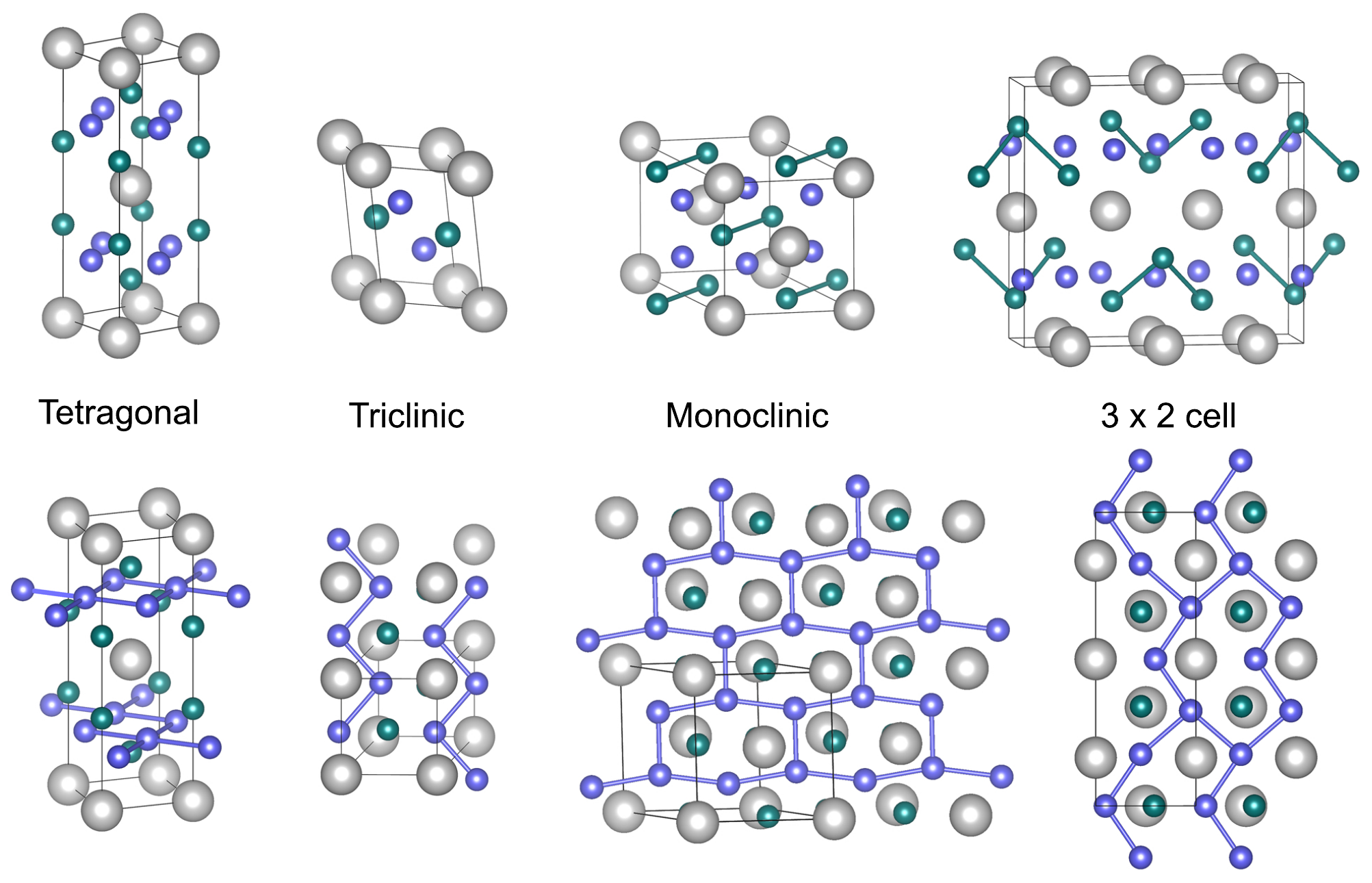}
 \caption{\label{fig:fig1} Structures of BaNi$_2$As$_2$ showing the network
of short As-As distances (top) and short Ni-Ni distances (bottom) in the ideal tetragonal, 
triclinic, based on relaxation of the structure of Sefat and co-workers \cite{Sefat_2009},
the monoclinic and the 3x2 structure.}
\end{figure*}

\begin{table*}[htpb]
\caption{\label{tab:table1}
Structural parameters after full relaxation and
the corresponding calculated total energy (meV/f.u.) relative to the tetragonal structure.
Results are for the PAW and LAPW methods and for the PBE and LDA functionals.
Note that for the 3$\times$2 structure, $b$ is the direction perpendicular to the NiAs planes.}
\begin{ruledtabular}
\begin{tabular}{cccc}
Crystal system&Tetragonal &Monoclinic & 3$\times$2 \\
\hline
Space group& $I4/mmm$ &$C2/m$ & $Pnnm$\\
Lattice parameter (\AA) & a = 4.139 & a = 5.711, b = 6.540 & a = 4.258, b = 11.533 \\
& c = 11.866  & c = 6.166, $\gamma$ = 117.68$^{\circ}$  &  c = 12.467  \\
Atomic positions & Ba: 0.0, 0.0, 0.0 &Ba: 0.0, 0.0, 0.0 &Ba1: 0.5, 0.5, 0.0; Ba2: 0.5, 0.5, 0.664\\
&Ni: 0.0, 0.5, 0.25  &Ni: 0.5, 0.5, 0.786&Ni1: 0.403, 0.244, 0.5; Ni2: 0.552, 0.251, 0.176\\
&As: 0.0, 0.0, 0.346&As: 0.812, 0.682, 0.5&As1: 0.546, 0.156, 0.661; As2: 0.418, 0.168, 0.0\\
PBE (PAW/LAPW) & 0 / 0 &-18 / -23 &-21 /-26 \\
LDA (PAW/LAPW) & 0 / 0 &-32 / -37 &-39 /-46 \\
\end{tabular}
\end{ruledtabular}
\end{table*}


This large hysteresis is different from the transition in BaFe$_2$As$_2$, and is also different from
typical charge density wave (CDW) instabilities.
It is noteworthy that there is a strong suppression of this structural phase transition with
P alloying in Ba(Ni$_{1-x}$P$_x$)$_2$As$_2$ with a concomitant enhancement of the superconducting critical temperature,
$T_c$
\cite{kudo,noda}.
However, the electronic specific heat is found to be similar for the distorted triclinic and ideal tetragonal
phases, different from the expectation for a CDW
\cite{kudo}.
Similarly, the transition can be suppressed by Sr substitution for Ba and Co substitution for Ni
\cite{Eckberg_2019,Eckberg_2018}.
Resistivity data for (Ba,Sr)Ni$_2$As$_2$ as a function
of strain shows a nematic-like lowering of rotational symmetry from tetragonal in the transition
\cite{Eckberg_2019}.
On the other hand, an additional periodicity was observed by Lee and co-workers in the vicinity of the
triclinic structure transition
\cite{Lee_2019a}.
Interestingly, this is initially incommensurate consisting of an approximate
tripling of the unit cell along one axis, but locks in at lower temperature.
The x-ray data is consistent with a coexistence of the distorted phase and the ideal tetragonal
structure down to low temperature
\cite{Lee_2019a}.
Recently, this distortion was further characterized using x-ray diffraction as a coexistence of three
distinct density waves with the conclusion that the observed
nematicity may be from antiphase domain walls rather than an intrinsic feature of the ground state
\cite{Lee_2021}.
Within a theory of nematic superconductivity this suggested superconductivity emerging from domain walls,
\cite{Lee_2021}.
However, this may be difficult to reconcile with the observed fully gapped state from specific heat
\cite{kurita}.
The strongly first order nature of the transition and the difficulty in establishing the detailed structure
suggests more complexity in the system than a soft phonon instability related to
Ni $d$ electrons. This may include competing metastable states, which might be consistent with
recent observations
\cite{Lee_2021}.
In any case, the detailed structure of the low temperature phase is clearly complex and not solved.

We started with first principles calculations examining the stability of the triclinic structure
proposed by Sefat and co-workers \cite{Sefat_2009}.
These calculations were done within density functional theory (DFT) using the standard generalized gradient
approximation of Perdew, Burke and Ernzerhof \cite{PBE}, as described below.
We did calculations relaxing the atomic positions fixing the lattice parameters to the
experimental triclinic values as given by Lee and co-workers
\cite{Lee_2019a}
and also as given by Sefat and co-workers
\cite{Sefat_2009}.

We find in both cases that while the triclinic structure, with zigzag chains of Ni atoms are maintained in the
relaxation, the energy differences with respect to the tetragonal are very small and in the case of the
structure of Lee and co-workers, the energy is higher than the tetragonal structure. In the case of the
structure of Sefat and co-workers, the energy is lower for the relaxed triclinic structure by less than
10 meV per formula unit, which is very small considering the first order transition at $\sim$ 130 K.
The structure highlighting the Ni-Ni chains is shown in comparison with the ideal tetragonal structure in
the first two panels of Fig. \ref{fig:fig1}.
We also tripled the the triclinic
unit cell as suggested by recent experimental results and relaxed after making various small
atomic displacements, but did not find any lower energy structure in these tripled cells.
This in combination with the experimental results discussed above suggest that the actual structure and
distortions in BaNi$_2$As$_2$ may be considerably more complex.
This in fact is the case, as discussed in the results below. In particular, BaNi$_2$As$_2$ is found to have
strong competing instabilities.
These are associated not directly with the electronic structure of the Ni $d$ electrons,
but are driven by As-As bonding.
This results in a complex interplay between the Ni $d$ electron physics, as manifested in transport,
and pnictogen bonding.

We searched for possible structures by starting with the tetragonal cell with various larger displacements of the
atoms, and then relaxing the structures.
We did this in a tripled conventional cell containing six formula units (denoted 3$\times$2 in the following).
No symmetry was imposed.
We used the projector augmented-wave (PAW) method \cite{VASP3} as implemented in \textsc{vasp}
\cite{VASP1,VASP2,VASP4} and the Perdew Burke Ernzerhof generalized gradient approximation (PBE GGA)
\cite{PBE}.
We used an energy cutoff of 550 eV and a Brillouin zone sampling with a
9$\times$9$\times$12 mesh.
We found two distinct low energy structures, with very similar energies.

The first is a monoclinic structure, with a unit cell containing a single formula unit,
as depicted in Fig. \ref{fig:fig1}.
As seen, it is very different from the triclinic structure. In particular, it is based
on a dimerization of the As ions. These displacements are coupled to the Ni layer, leading to
an anisotropic arrangement of the Ni atoms to form chain-like structures as shown.
This leads to a large anisotropy in the conductivity, as discussed below.
The space group and structural parameters are given in Table \ref{tab:table1}.
The basic structural motif in tetragonal BaNi$_2$As$_2$ is square planar sheets of Ni atoms, which
are tetrahedrally coordinated by As above and below the sheets.

In the PBE relaxed tetragonal structure
each Ni has four Ni at 2.93 \AA, and
each As has one As at 3.65 \AA, and four at 3.71 \AA.
In the triclinic structure with relaxed atomic positions
and lattice parameters constrained to those of Sefat and co-workers
\cite{Sefat_2009}, the Ni-Ni distances become
two short distances of 2.77 \AA, and 2.78 \AA, and two longer distances of 3.13 \AA,
leading to zigzag chains. The As distances are also changed. The shortest As-As distance
becomes 3.55 \AA, while two other As come closer at $\sim$3.58 \AA, and two move further to $\sim$3.81 \AA.
In contrast, the monoclinic structure that we find has a much stronger
distortion of the As positions.
It has a short As-As distance of 3.23 \AA, with the other four nearby As in the range
3.69 \AA -- 3.88 \AA.
Thus the As form dimers as shown in Fig. \ref{fig:fig1}.
This drives a distortion of the Ni sublattice, so that the four shortest Ni-Ni distances become
2.64 \AA, 2.89 \AA, 2.89 \AA, and 3.53 \AA.
As seen in Fig. \ref{fig:fig1},
this again leads to a loss of four fold rotational invariance in the Ni layers, but with a more complicated
pattern that the zigzag chains.

\begin{figure}[htbp]
 \centering
 \includegraphics[width=1.0\columnwidth]{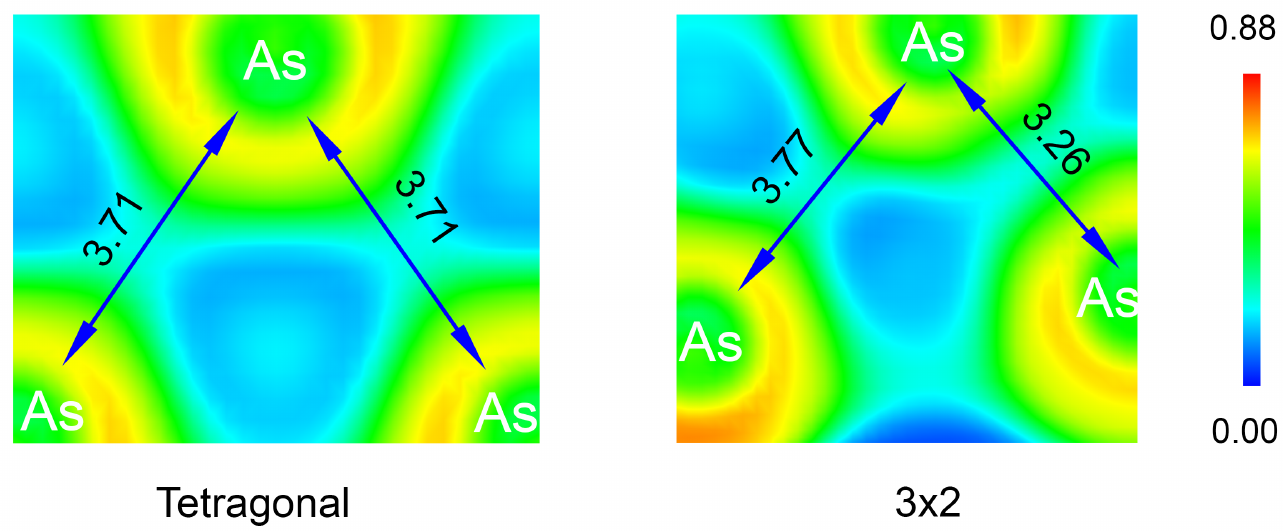}
 \caption{\label{fig:elf} Comparison of the ELF for the ideal tetragonal and 3x2 structures showing
a plane containing an As trimer. The bond lengths are as indicated.}
\end{figure}

\begin{figure}[htbp]
 \centering
 \includegraphics[width=0.9\columnwidth]{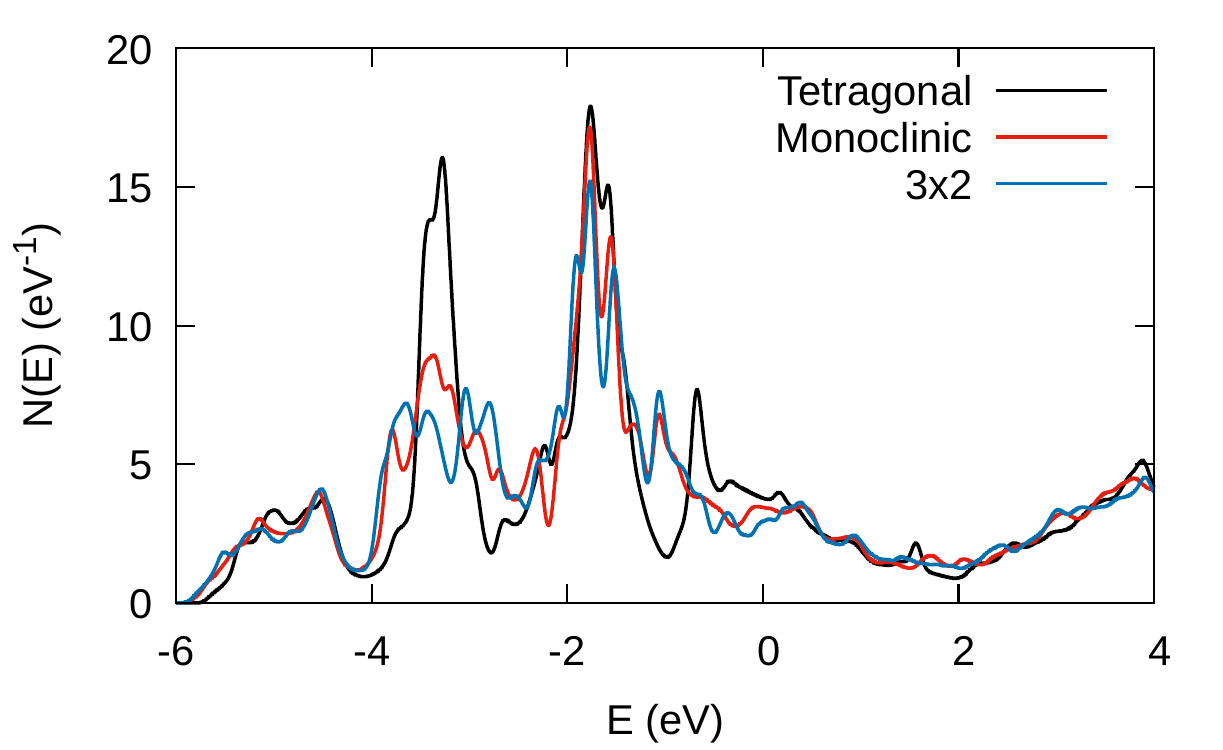}
 \caption{\label{fig:dos} Calculated electronic density of states in the tetragonal,
monoclinic and 3$\times$2 structures. The Fermi level is at zero.}
\end{figure}

\begin{figure}[htbp]
 \centering
 \includegraphics[width=0.75\columnwidth]{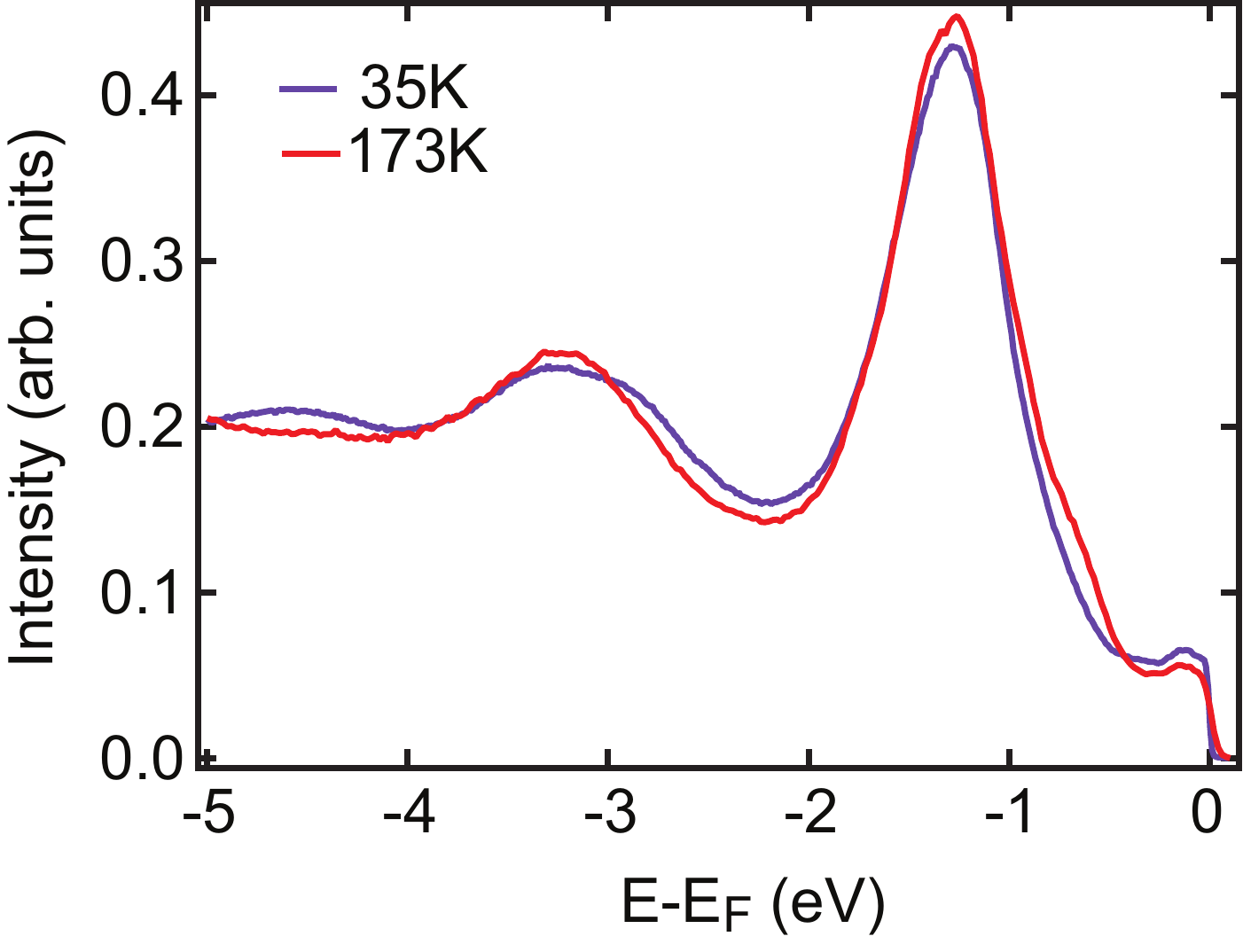}
 \caption{\label{fig:aipes} Angle integrated photoemission spectra at 173 K and 35 K, i.e. in the
tetragonal and low temperature structures.}
\end{figure}

\begin{figure}[htbp]
 \centering
 \includegraphics[width=1.0\columnwidth]{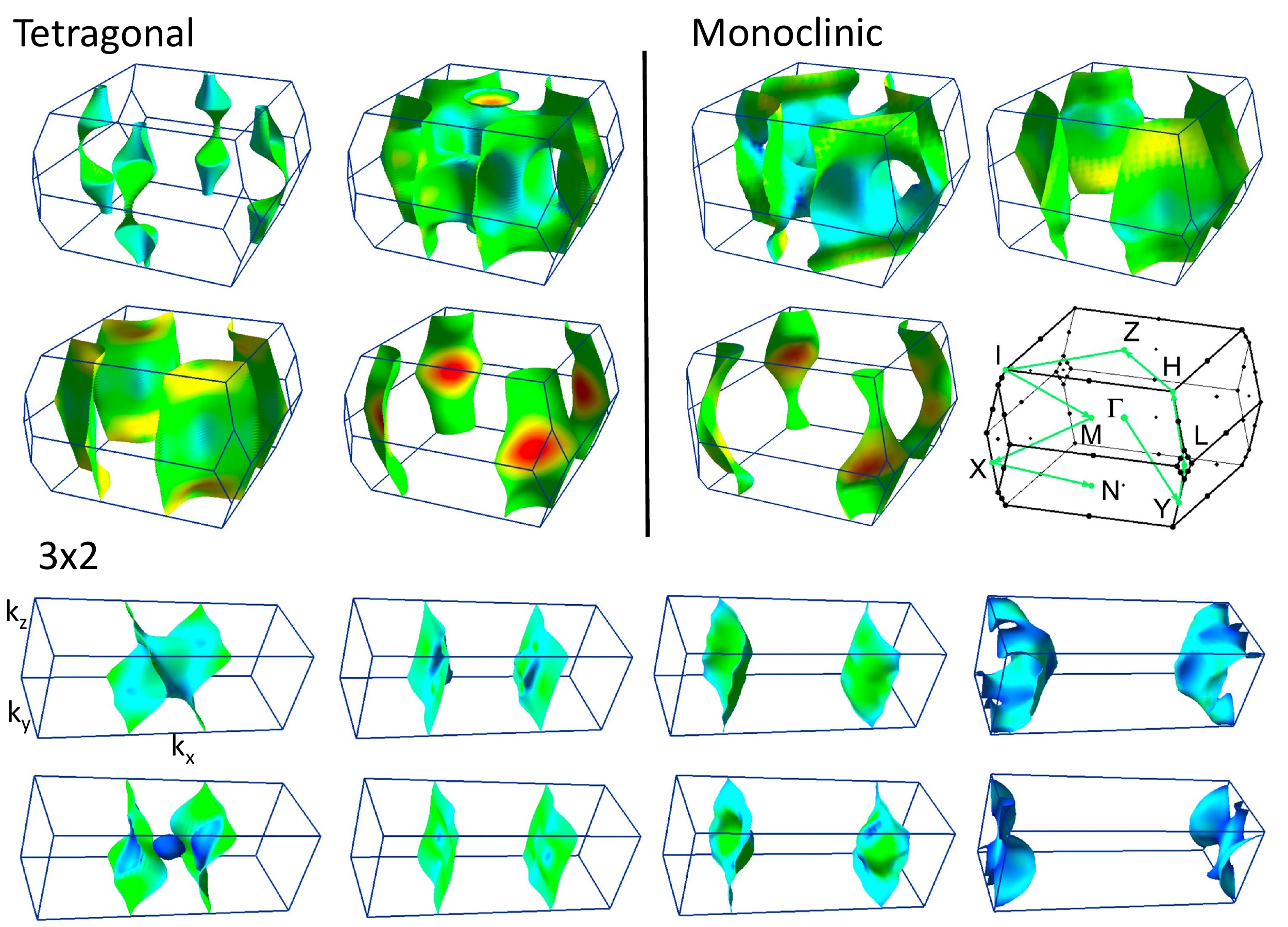}
 \caption{\label{fig:fermi} Fermi surfaces colored by velocity in the tetragonal,
monoclinic and 3$\times$2 structures. Also shown is the Brillouin zone for the monoclinic structure.
Blue is low velocity and red is high.}
\end{figure}

\begin{table}[tb]
\caption{\label{tab:table2}
Total density of states (electrons/f.u.) at Fermi level N(E$_F$) and $\sigma$/$\tau$ ($10^{20}$/$\Omega$ms)
for the tetragonal, monoclinic and 3$\times$2 structures. For the monoclinic structure the eigenvalues of the
$\sigma$/$\tau$ tensor are given. For the 3$\times$2 structure, the first two values are the in plane
short and long directions and the final value is the $c$-axis direction perpendicular to the NiAs layers.}
\begin{ruledtabular}
\begin{tabular}{ccccc}
Phase& N(E$_F$) & $\sigma_x$/$\tau$ & $\sigma_y$/ $\tau$ &$\sigma_z$/$\tau$\\
\hline
Tetragonal & 3.78 & 4.98 & 4.98 & 1.65 \\
Monoclinic & 3.45 & 4.31 & 3.83 & 1.30 \\
3$\times$2 & 2.94 & 2.08 & 0.56 & 0.57 \\     
\end{tabular}
\end{ruledtabular}
\end{table}

The second low energy structure (the 3$\times$2 structure) has an orthorhombic cell corresponding to the
conventional cell tripled along an in-plane lattice direction.
It has spacegroup $Pnnm$.
Similar to the monoclinic structure, it is based on strong distortion of the As framework, which
drives a distortion of the Ni sublattice, as shown in Fig. \ref{fig:fig1}.
Structural parameters are in Table \ref{tab:table1}.
The nature of the distortion is, however, very different from the monoclinic structure.
It may be described as a trimerization of the As yielding bent As$_3$ units instead
of the dimers characterizing the monoclinic structure.
These are stacked oppositely along the
$c$-axis direction, leading to a six formula unit cell.
The As-As bond lengths in the trimers are 3.26 \AA. The other As-As distances
range from 3.62 \AA to 3.94 \AA.
Fig. \ref{fig:elf}
shows a comparison of the electron localization function (ELF)
\cite{elf} in the plane of one of the trimers with
a similar plane in the ideal tetragonal structure.
The ELF shows bonding interactions between the As atoms making up a trimer.
Similar to the monoclinic structure this distortion of the As sublattice drives a reconstruction of the Ni 
sublattice, again leading to a clear visible loss of four fold rotational symmetry. The Ni-Ni distances,
all equal in the ideal tetragonal structure, are in the range 2.65 \AA--3.53 \AA, demonstrating a very
large distortion.

We checked these results by further relaxing internal coordinates and calculating the energies using
the linearized augmented planewave method as implemented in WIEN2k
\cite{WIEN2k}, but did not find any significant differences.
We also did calculations using the local density approximation (LDA), but again find similar structures.
Details are in Table \ref{tab:table1}.
The 3$\times$2 structure has the lowest energy.
Importantly, it and the monoclinic have similar
energies, and these significantly lower than either the ideal tetragonal or the triclinic structures.

We now turn to the electronic structure.
The distortions strongly affect the electronic structure, including near the Fermi level,
where the states are mainly Ni $d$ derived.
The electronic densities of states (DOS) of the two low energy structures
are compared with the ideal tetragonal structure in Fig. \ref{fig:dos}.
As seen there is a considerable reconstruction of the DOS, including the region near the Fermi level, $E_F$
where the states are primarily Ni $d$ derived. This is reflected in the values at the Fermi level, $N(E_F)$
as given in Table \ref{tab:table2}.
There are also substantial changes in the DOS near $\sim$ -1 eV and $\sim$ -3.4 eV.

We additionally did angle integrated photoemission spectroscopy to probe the DOS
in the tetragonal and lower symmetry distrorted structure. The result is as shown in Fig. \ref{fig:aipes}.
High-quality single crystals of BaNi$_2$As$_2$ were synthesized using the self-flux method.
Angle-integrated photoemission spectroscopy were performed using a DA30L analyzer and helium discharge
lamp with a typical energy resolution of 12 meV.
The samples were cleaved in-situ with a base pressure below 5×10$^{−11}$ torr. 
A prominent peak appears in the measured DOS at approximately -1.3 eV,
while two broader peaks are located at $\sim$-3 eV and $\sim$-4.5 eV.
The overall structure is reasonably consistent with the calculated DOS.
When temperature is lowered across the structural transition,
the $\sim$-3 eV peak appears to broaden. This is also consistent with the overall trend
in the calculated DOS from the tetragonal phase to the lower symmetry phase.

The Fermi surfaces are stromgly affected by the transition. 
The tetragonal structure has four sheets of Fermi surface as seen. The monoclinic has only three sheets,
with a noticeable anisotropy. This is quantified by the transport function $\sigma/\tau$
as given in Table \ref{tab:table2}.
This was obtained using the BoltzTraP code \cite{Madsen_2006}.
Larger changes in the Fermi surfaces and transport function are found in the 3$\times$2 structure.
The zone folding associated with the formation of the six formula unit cell leads to eight bands crossing
the Fermi energy.
Importantly, while the in-plane conductivity is reduced in
both directions, it is very strongly reduced in direction of the unit tripling.
This leads to a prediction of a very strong transport anisotropy.

The characteristic of the two distortions that we find is a formation of dimers or trimers of As. 
The results imply rebonding of the As as an important ingredient in understanding the structural
distortions of BaNi$_2$As$_2$.
The formation of the electronic structure of $AB_2X_2$ ThCr$_2$Si$_2$-type materials is known to depend on both 
$B$-$X$ and $X$-$X$ bonding, which can vary between compounds
\cite{hoffmann}.
Here we find that, as in the electronic structure,
the structural stability and distortions of BaNi$_2$As$_2$
are governed by both Ni-As and As-As interactions.
This illustrates a fundamental difference between the physics of correlated transition metal oxides
and correlated pnictide metals.
Oxygen anions do not normally bond in transition metal oxides compounds. This is a consequence of the
relatively small effective size of O$^{2-}$ and the high electronegativity of O. As a result structural
instabilities in oxides are normally a consequence of instabilities associated with steric
effects, for example mismatches of ionic radii in perovskites and $d$ electrons physics, particularly
Fermi surface instabilities, orbital orderings and the interplay of structure with Mott physics.
However, other chalcogenides, particularly tellurides can have structures driven by chalcogen-chalcogen
bonding. This is because of the lower electronegativity and larger size of Te anions, as seen for example
in the first order structural phase transition of IrTe$_2$ 
\cite{cao-irte2,li-irte2}.
Here we find an interplay between the $d$ electron physics and the structure with distortions related to
As bonding.
The present results show the importance of these effects in the physics of BaNi$_2$As$_2$.

To summarize, we find that the structural instability of BaNi$_2$As$_2$ is related to rebonding of the As anions,
which then drives reconstruction of the electronic structure. We find low energy structures related to the formation
of As dimers and trimers. Other structures with more complex bonding arrangements of the As
may occur. It is likely that the competition between structures underlies the complex structural
and electronic behavior observed in BaNi$_2$As$_2$.
There is a strong interplay of these bonding arrangements with the Ni derived electronic structure,
including strong changes in the Fermiology and transport function. This may provide novel routes to modify and
control the electronic properties of correlated metals and exploring the physics of such systems.

\begin{acknowledgments}

Theory work at the University of Missouri
was supported by the Department of Energy, Basic Energy Sciences, Award DE-SC0019114.
Crystal synthesis at Rice University was supported by US DOE DE-SC0012311 and by the Robert A. Welch Foundation
under Grant No. C-1839 (P.D.)..

\end{acknowledgments}

\bibliography{BaNi2As2}

\begin{thebibliography}{30}%
\makeatletter
\providecommand \@ifxundefined [1]{%
 \@ifx{#1\undefined}
}%
\providecommand \@ifnum [1]{%
 \ifnum #1\expandafter \@firstoftwo
 \else \expandafter \@secondoftwo
 \fi
}%
\providecommand \@ifx [1]{%
 \ifx #1\expandafter \@firstoftwo
 \else \expandafter \@secondoftwo
 \fi
}%
\providecommand \natexlab [1]{#1}%
\providecommand \enquote  [1]{``#1''}%
\providecommand \bibnamefont  [1]{#1}%
\providecommand \bibfnamefont [1]{#1}%
\providecommand \citenamefont [1]{#1}%
\providecommand \href@noop [0]{\@secondoftwo}%
\providecommand \href [0]{\begingroup \@sanitize@url \@href}%
\providecommand \@href[1]{\@@startlink{#1}\@@href}%
\providecommand \@@href[1]{\endgroup#1\@@endlink}%
\providecommand \@sanitize@url [0]{\catcode `\\12\catcode `\$12\catcode
  `\&12\catcode `\#12\catcode `\^12\catcode `\_12\catcode `\%12\relax}%
\providecommand \@@startlink[1]{}%
\providecommand \@@endlink[0]{}%
\providecommand \url  [0]{\begingroup\@sanitize@url \@url }%
\providecommand \@url [1]{\endgroup\@href {#1}{\urlprefix }}%
\providecommand \urlprefix  [0]{URL }%
\providecommand \Eprint [0]{\href }%
\providecommand \doibase [0]{http://dx.doi.org/}%
\providecommand \selectlanguage [0]{\@gobble}%
\providecommand \bibinfo  [0]{\@secondoftwo}%
\providecommand \bibfield  [0]{\@secondoftwo}%
\providecommand \translation [1]{[#1]}%
\providecommand \BibitemOpen [0]{}%
\providecommand \bibitemStop [0]{}%
\providecommand \bibitemNoStop [0]{.\EOS\space}%
\providecommand \EOS [0]{\spacefactor3000\relax}%
\providecommand \BibitemShut  [1]{\csname bibitem#1\endcsname}%
\let\auto@bib@innerbib\@empty
\bibitem [{\citenamefont {Kamihara}\ \emph {et~al.}(2008)\citenamefont
  {Kamihara}, \citenamefont {Watanabe}, \citenamefont {Hirano},\ and\
  \citenamefont {Hosono}}]{Kamihara_2008}%
  \BibitemOpen
  \bibfield  {author} {\bibinfo {author} {\bibfnamefont {Y.}~\bibnamefont
  {Kamihara}}, \bibinfo {author} {\bibfnamefont {T.}~\bibnamefont {Watanabe}},
  \bibinfo {author} {\bibfnamefont {M.}~\bibnamefont {Hirano}}, \ and\ \bibinfo
  {author} {\bibfnamefont {H.}~\bibnamefont {Hosono}},\ }\href {\doibase
  10.1021/ja800073m} {\bibfield  {journal} {\bibinfo  {journal} {J. Am. Chem.
  Soc.}\ }\textbf {\bibinfo {volume} {130}},\ \bibinfo {pages} {3296} (\bibinfo
  {year} {2008})}\BibitemShut {NoStop}%
\bibitem [{\citenamefont {Johnston}(2010)}]{johnston}%
  \BibitemOpen
  \bibfield  {author} {\bibinfo {author} {\bibfnamefont {D.~C.}\ \bibnamefont
  {Johnston}},\ }\href {https://doi.org/10.1080/00018732.2010.513480}
  {\bibfield  {journal} {\bibinfo  {journal} {Adv. Phys.}\ }\textbf {\bibinfo
  {volume} {59}},\ \bibinfo {pages} {803} (\bibinfo {year} {2010})}\BibitemShut
  {NoStop}%
\bibitem [{\citenamefont {Mazin}\ \emph {et~al.}(2008)\citenamefont {Mazin},
  \citenamefont {Singh}, \citenamefont {Johannes},\ and\ \citenamefont
  {Du}}]{Mazin_2008}%
  \BibitemOpen
  \bibfield  {author} {\bibinfo {author} {\bibfnamefont {I.~I.}\ \bibnamefont
  {Mazin}}, \bibinfo {author} {\bibfnamefont {D.~J.}\ \bibnamefont {Singh}},
  \bibinfo {author} {\bibfnamefont {M.~D.}\ \bibnamefont {Johannes}}, \ and\
  \bibinfo {author} {\bibfnamefont {M.~H.}\ \bibnamefont {Du}},\ }\href@noop {}
  {\bibfield  {journal} {\bibinfo  {journal} {Phys. Rev. Lett.}\ }\textbf
  {\bibinfo {volume} {101}},\ \bibinfo {pages} {057003} (\bibinfo {year}
  {2008})}\BibitemShut {NoStop}%
\bibitem [{\citenamefont {Dai}(2015)}]{dai}%
  \BibitemOpen
  \bibfield  {author} {\bibinfo {author} {\bibfnamefont {P.}~\bibnamefont
  {Dai}},\ }\href {https://doi.org/10.1103/RevModPhys.87.855} {\bibfield
  {journal} {\bibinfo  {journal} {Rev. Mod. Phys.}\ }\textbf {\bibinfo {volume}
  {87}},\ \bibinfo {pages} {855} (\bibinfo {year} {2015})}\BibitemShut
  {NoStop}%
\bibitem [{\citenamefont {Si}\ \emph {et~al.}(2016)\citenamefont {Si},
  \citenamefont {Yu},\ and\ \citenamefont {Abrahams}}]{si}%
  \BibitemOpen
  \bibfield  {author} {\bibinfo {author} {\bibfnamefont {Q.}~\bibnamefont
  {Si}}, \bibinfo {author} {\bibfnamefont {R.}~\bibnamefont {Yu}}, \ and\
  \bibinfo {author} {\bibfnamefont {E.}~\bibnamefont {Abrahams}},\ }\href
  {https://doi.org/10.1038/natrevmats.2016.17} {\bibfield  {journal} {\bibinfo
  {journal} {Nature Reviews Materials}\ }\textbf {\bibinfo {volume} {1}},\
  \bibinfo {pages} {16017} (\bibinfo {year} {2016})}\BibitemShut {NoStop}%
\bibitem [{\citenamefont {Worasaran}\ \emph {et~al.}(2021)\citenamefont
  {Worasaran}, \citenamefont {Ikeda}, \citenamefont {Palmstrom}, \citenamefont
  {Straquadine}, \citenamefont {Kivelson},\ and\ \citenamefont
  {Fisher}}]{worasaran}%
  \BibitemOpen
  \bibfield  {author} {\bibinfo {author} {\bibfnamefont {T.}~\bibnamefont
  {Worasaran}}, \bibinfo {author} {\bibfnamefont {M.~S.}\ \bibnamefont
  {Ikeda}}, \bibinfo {author} {\bibfnamefont {J.~C.}\ \bibnamefont
  {Palmstrom}}, \bibinfo {author} {\bibfnamefont {J.~A.~W.}\ \bibnamefont
  {Straquadine}}, \bibinfo {author} {\bibfnamefont {S.~A.}\ \bibnamefont
  {Kivelson}}, \ and\ \bibinfo {author} {\bibfnamefont {I.~R.}\ \bibnamefont
  {Fisher}},\ }\href {https://doi.org/10.1126/science.abb9280} {\bibfield
  {journal} {\bibinfo  {journal} {Science}\ }\textbf {\bibinfo {volume}
  {372}},\ \bibinfo {pages} {973} (\bibinfo {year} {2021})}\BibitemShut
  {NoStop}%
\bibitem [{\citenamefont {Ronning}\ \emph {et~al.}(2008)\citenamefont
  {Ronning}, \citenamefont {Kurita}, \citenamefont {Bauer}, \citenamefont
  {Scott}, \citenamefont {Park}, \citenamefont {Klimczuk}, \citenamefont
  {Movshovich},\ and\ \citenamefont {Thompson}}]{Ronning_2008}%
  \BibitemOpen
  \bibfield  {author} {\bibinfo {author} {\bibfnamefont {F.}~\bibnamefont
  {Ronning}}, \bibinfo {author} {\bibfnamefont {N.}~\bibnamefont {Kurita}},
  \bibinfo {author} {\bibfnamefont {E.~D.}\ \bibnamefont {Bauer}}, \bibinfo
  {author} {\bibfnamefont {B.~L.}\ \bibnamefont {Scott}}, \bibinfo {author}
  {\bibfnamefont {T.}~\bibnamefont {Park}}, \bibinfo {author} {\bibfnamefont
  {T.}~\bibnamefont {Klimczuk}}, \bibinfo {author} {\bibfnamefont
  {R.}~\bibnamefont {Movshovich}}, \ and\ \bibinfo {author} {\bibfnamefont
  {J.~D.}\ \bibnamefont {Thompson}},\ }\href {\doibase
  10.1088/0953-8984/20/34/342203} {\bibfield  {journal} {\bibinfo  {journal}
  {J. Phys.: Condens. Matter}\ }\textbf {\bibinfo {volume} {20}},\ \bibinfo
  {pages} {342203} (\bibinfo {year} {2008})}\BibitemShut {NoStop}%
\bibitem [{\citenamefont {Pfisterer}\ and\ \citenamefont
  {Nagorsen}(1980)}]{pfisterer}%
  \BibitemOpen
  \bibfield  {author} {\bibinfo {author} {\bibfnamefont {M.}~\bibnamefont
  {Pfisterer}}\ and\ \bibinfo {author} {\bibfnamefont {G.}~\bibnamefont
  {Nagorsen}},\ }\href {https://doi.org/10.1515/znb-1980-0611} {\bibfield
  {journal} {\bibinfo  {journal} {Z. Naturforsch.}\ }\textbf {\bibinfo {volume}
  {86b}},\ \bibinfo {pages} {703} (\bibinfo {year} {1980})}\BibitemShut
  {NoStop}%
\bibitem [{\citenamefont {Kurita}\ \emph {et~al.}(2009)\citenamefont {Kurita},
  \citenamefont {Ronning}, \citenamefont {Tokiwa}, \citenamefont {Bauer},
  \citenamefont {Subedi}, \citenamefont {Singh}, \citenamefont {Thompson},\
  and\ \citenamefont {Movshovich}}]{kurita}%
  \BibitemOpen
  \bibfield  {author} {\bibinfo {author} {\bibfnamefont {N.}~\bibnamefont
  {Kurita}}, \bibinfo {author} {\bibfnamefont {F.}~\bibnamefont {Ronning}},
  \bibinfo {author} {\bibfnamefont {Y.}~\bibnamefont {Tokiwa}}, \bibinfo
  {author} {\bibfnamefont {E.~D.}\ \bibnamefont {Bauer}}, \bibinfo {author}
  {\bibfnamefont {A.}~\bibnamefont {Subedi}}, \bibinfo {author} {\bibfnamefont
  {D.~J.}\ \bibnamefont {Singh}}, \bibinfo {author} {\bibfnamefont {J.~D.}\
  \bibnamefont {Thompson}}, \ and\ \bibinfo {author} {\bibfnamefont
  {R.}~\bibnamefont {Movshovich}},\ }\href
  {https://doi.org/10.1103/PhysRevLett.102.147004} {\bibfield  {journal}
  {\bibinfo  {journal} {Phys. Rev. Lett.}\ }\textbf {\bibinfo {volume} {102}},\
  \bibinfo {pages} {147004} (\bibinfo {year} {2009})}\BibitemShut {NoStop}%
\bibitem [{\citenamefont {Yamakawa}\ \emph {et~al.}(2013)\citenamefont
  {Yamakawa}, \citenamefont {Onari},\ and\ \citenamefont {Kontani}}]{yamakawa}%
  \BibitemOpen
  \bibfield  {author} {\bibinfo {author} {\bibfnamefont {Y.}~\bibnamefont
  {Yamakawa}}, \bibinfo {author} {\bibfnamefont {S.}~\bibnamefont {Onari}}, \
  and\ \bibinfo {author} {\bibfnamefont {H.}~\bibnamefont {Kontani}},\ }\href
  {https://dx.doi.org/10.7566/JPSJ.82.094704} {\bibfield  {journal} {\bibinfo
  {journal} {J. Phys. Soc. Jpn.}\ }\textbf {\bibinfo {volume} {82}},\ \bibinfo
  {pages} {093704} (\bibinfo {year} {2013})}\BibitemShut {NoStop}%
\bibitem [{\citenamefont {Sefat}\ \emph {et~al.}(2009)\citenamefont {Sefat},
  \citenamefont {{McGuire}}, \citenamefont {Jin}, \citenamefont {Sales},
  \citenamefont {Mandrus}, \citenamefont {Ronning}, \citenamefont {Bauer},\
  and\ \citenamefont {Mozharivskyj}}]{Sefat_2009}%
  \BibitemOpen
  \bibfield  {author} {\bibinfo {author} {\bibfnamefont {A.~S.}\ \bibnamefont
  {Sefat}}, \bibinfo {author} {\bibfnamefont {M.~A.}\ \bibnamefont
  {{McGuire}}}, \bibinfo {author} {\bibfnamefont {R.}~\bibnamefont {Jin}},
  \bibinfo {author} {\bibfnamefont {B.~C.}\ \bibnamefont {Sales}}, \bibinfo
  {author} {\bibfnamefont {D.}~\bibnamefont {Mandrus}}, \bibinfo {author}
  {\bibfnamefont {F.}~\bibnamefont {Ronning}}, \bibinfo {author} {\bibfnamefont
  {E.~D.}\ \bibnamefont {Bauer}}, \ and\ \bibinfo {author} {\bibfnamefont
  {Y.}~\bibnamefont {Mozharivskyj}},\ }\href
  {https://doi.org/10.1103/PhysRevB.79.094508} {\bibfield  {journal} {\bibinfo
  {journal} {Phys. Rev. B}\ }\textbf {\bibinfo {volume} {79}},\ \bibinfo
  {pages} {094508} (\bibinfo {year} {2009})}\BibitemShut {NoStop}%
\bibitem [{\citenamefont {Zhou}\ \emph {et~al.}(2011)\citenamefont {Zhou},
  \citenamefont {Xu}, \citenamefont {Zhang}, \citenamefont {Xu}, \citenamefont
  {He}, \citenamefont {Yang}, \citenamefont {Chen}, \citenamefont {Xie},
  \citenamefont {Cui}, \citenamefont {Arita}, \citenamefont {Shimada},
  \citenamefont {Namatame}, \citenamefont {Taniguchi}, \citenamefont {Dai},\
  and\ \citenamefont {Feng}}]{Zhou_2011}%
  \BibitemOpen
  \bibfield  {author} {\bibinfo {author} {\bibfnamefont {B.}~\bibnamefont
  {Zhou}}, \bibinfo {author} {\bibfnamefont {M.}~\bibnamefont {Xu}}, \bibinfo
  {author} {\bibfnamefont {Y.}~\bibnamefont {Zhang}}, \bibinfo {author}
  {\bibfnamefont {G.}~\bibnamefont {Xu}}, \bibinfo {author} {\bibfnamefont
  {C.}~\bibnamefont {He}}, \bibinfo {author} {\bibfnamefont {L.~X.}\
  \bibnamefont {Yang}}, \bibinfo {author} {\bibfnamefont {F.}~\bibnamefont
  {Chen}}, \bibinfo {author} {\bibfnamefont {B.~P.}\ \bibnamefont {Xie}},
  \bibinfo {author} {\bibfnamefont {X.-Y.}\ \bibnamefont {Cui}}, \bibinfo
  {author} {\bibfnamefont {M.}~\bibnamefont {Arita}}, \bibinfo {author}
  {\bibfnamefont {K.}~\bibnamefont {Shimada}}, \bibinfo {author} {\bibfnamefont
  {H.}~\bibnamefont {Namatame}}, \bibinfo {author} {\bibfnamefont
  {M.}~\bibnamefont {Taniguchi}}, \bibinfo {author} {\bibfnamefont
  {X.}~\bibnamefont {Dai}}, \ and\ \bibinfo {author} {\bibfnamefont {D.~L.}\
  \bibnamefont {Feng}},\ }\href {https://doi.org/10.1103%2Fphysrevb.83.035110}
  {\bibfield  {journal} {\bibinfo  {journal} {Phys. Rev. B}\ }\textbf {\bibinfo
  {volume} {83}},\ \bibinfo {pages} {035110} (\bibinfo {year}
  {2011})}\BibitemShut {NoStop}%
\bibitem [{\citenamefont {Kothapalli}\ \emph {et~al.}(2010)\citenamefont
  {Kothapalli}, \citenamefont {Ronning}, \citenamefont {Bauer}, \citenamefont
  {Schultz},\ and\ \citenamefont {Nakotte}}]{kothapalli}%
  \BibitemOpen
  \bibfield  {author} {\bibinfo {author} {\bibfnamefont {K.}~\bibnamefont
  {Kothapalli}}, \bibinfo {author} {\bibfnamefont {F.}~\bibnamefont {Ronning}},
  \bibinfo {author} {\bibfnamefont {E.~D.}\ \bibnamefont {Bauer}}, \bibinfo
  {author} {\bibfnamefont {A.~J.}\ \bibnamefont {Schultz}}, \ and\ \bibinfo
  {author} {\bibfnamefont {H.}~\bibnamefont {Nakotte}},\ }\href
  {https://doi.org/10.1088/1742-6596/251/1/012010} {\bibfield  {journal}
  {\bibinfo  {journal} {J. Phys.: Conf. Ser.}\ }\textbf {\bibinfo {volume}
  {251}},\ \bibinfo {pages} {012010} (\bibinfo {year} {2010})}\BibitemShut
  {NoStop}%
\bibitem [{\citenamefont {Kudo}\ \emph {et~al.}(2012)\citenamefont {Kudo},
  \citenamefont {Takasuga}, \citenamefont {Okamoto}, \citenamefont {Hiroi},\
  and\ \citenamefont {Nohara}}]{kudo}%
  \BibitemOpen
  \bibfield  {author} {\bibinfo {author} {\bibfnamefont {K.}~\bibnamefont
  {Kudo}}, \bibinfo {author} {\bibfnamefont {M.}~\bibnamefont {Takasuga}},
  \bibinfo {author} {\bibfnamefont {Y.}~\bibnamefont {Okamoto}}, \bibinfo
  {author} {\bibfnamefont {Z.}~\bibnamefont {Hiroi}}, \ and\ \bibinfo {author}
  {\bibfnamefont {M.}~\bibnamefont {Nohara}},\ }\href
  {https://doi.org/10.1103/PhysRevLett.109.097002} {\bibfield  {journal}
  {\bibinfo  {journal} {Phys. Rev. Lett.}\ }\textbf {\bibinfo {volume} {109}},\
  \bibinfo {pages} {097002} (\bibinfo {year} {2012})}\BibitemShut {NoStop}%
\bibitem [{\citenamefont {Noda}\ \emph {et~al.}(2017)\citenamefont {Noda},
  \citenamefont {Kudo}, \citenamefont {Takasuga}, \citenamefont {Nohara},
  \citenamefont {Sugimoto}, \citenamefont {Ootsuki}, \citenamefont {Kobayashi},
  \citenamefont {Horiba}, \citenamefont {Ono}, \citenamefont {Kumigashira},
  \citenamefont {Fujimori}, \citenamefont {Saini},\ and\ \citenamefont
  {Mizokawa}}]{noda}%
  \BibitemOpen
  \bibfield  {author} {\bibinfo {author} {\bibfnamefont {T.}~\bibnamefont
  {Noda}}, \bibinfo {author} {\bibfnamefont {K.}~\bibnamefont {Kudo}}, \bibinfo
  {author} {\bibfnamefont {M.}~\bibnamefont {Takasuga}}, \bibinfo {author}
  {\bibfnamefont {M.}~\bibnamefont {Nohara}}, \bibinfo {author} {\bibfnamefont
  {T.}~\bibnamefont {Sugimoto}}, \bibinfo {author} {\bibfnamefont
  {D.}~\bibnamefont {Ootsuki}}, \bibinfo {author} {\bibfnamefont
  {M.}~\bibnamefont {Kobayashi}}, \bibinfo {author} {\bibfnamefont
  {K.}~\bibnamefont {Horiba}}, \bibinfo {author} {\bibfnamefont
  {K.}~\bibnamefont {Ono}}, \bibinfo {author} {\bibfnamefont {H.}~\bibnamefont
  {Kumigashira}}, \bibinfo {author} {\bibfnamefont {A.}~\bibnamefont
  {Fujimori}}, \bibinfo {author} {\bibfnamefont {N.~L.}\ \bibnamefont {Saini}},
  \ and\ \bibinfo {author} {\bibfnamefont {T.}~\bibnamefont {Mizokawa}},\
  }\href {https://doi.org/10.7566/JPSJ.86.064708} {\bibfield  {journal}
  {\bibinfo  {journal} {J. Phys. Soc. Jpn.}\ }\textbf {\bibinfo {volume}
  {86}},\ \bibinfo {pages} {064708} (\bibinfo {year} {2017})}\BibitemShut
  {NoStop}%
\bibitem [{\citenamefont {Eckberg}\ \emph {et~al.}(2019)\citenamefont
  {Eckberg}, \citenamefont {Campbell}, \citenamefont {Metz}, \citenamefont
  {Collini}, \citenamefont {Hodovanets}, \citenamefont {Drye}, \citenamefont
  {Zavalij}, \citenamefont {Christensen}, \citenamefont {Fernandes},
  \citenamefont {Lee}, \citenamefont {Abbamonte}, \citenamefont {Lynn},\ and\
  \citenamefont {Paglione}}]{Eckberg_2019}%
  \BibitemOpen
  \bibfield  {author} {\bibinfo {author} {\bibfnamefont {C.}~\bibnamefont
  {Eckberg}}, \bibinfo {author} {\bibfnamefont {D.~J.}\ \bibnamefont
  {Campbell}}, \bibinfo {author} {\bibfnamefont {T.}~\bibnamefont {Metz}},
  \bibinfo {author} {\bibfnamefont {J.}~\bibnamefont {Collini}}, \bibinfo
  {author} {\bibfnamefont {H.}~\bibnamefont {Hodovanets}}, \bibinfo {author}
  {\bibfnamefont {T.}~\bibnamefont {Drye}}, \bibinfo {author} {\bibfnamefont
  {P.}~\bibnamefont {Zavalij}}, \bibinfo {author} {\bibfnamefont {M.~H.}\
  \bibnamefont {Christensen}}, \bibinfo {author} {\bibfnamefont {R.~M.}\
  \bibnamefont {Fernandes}}, \bibinfo {author} {\bibfnamefont {S.}~\bibnamefont
  {Lee}}, \bibinfo {author} {\bibfnamefont {P.}~\bibnamefont {Abbamonte}},
  \bibinfo {author} {\bibfnamefont {J.~W.}\ \bibnamefont {Lynn}}, \ and\
  \bibinfo {author} {\bibfnamefont {J.}~\bibnamefont {Paglione}},\ }\href@noop
  {} {\bibfield  {journal} {\bibinfo  {journal} {Nature Physics}\ }\textbf
  {\bibinfo {volume} {16}},\ \bibinfo {pages} {346} (\bibinfo {year}
  {2019})}\BibitemShut {NoStop}%
\bibitem [{\citenamefont {Eckberg}\ \emph {et~al.}(2018)\citenamefont
  {Eckberg}, \citenamefont {Wang}, \citenamefont {Hodovanets}, \citenamefont
  {Kim}, \citenamefont {Campbell}, \citenamefont {Zavalij}, \citenamefont
  {Piccoli},\ and\ \citenamefont {Paglione}}]{Eckberg_2018}%
  \BibitemOpen
  \bibfield  {author} {\bibinfo {author} {\bibfnamefont {C.}~\bibnamefont
  {Eckberg}}, \bibinfo {author} {\bibfnamefont {L.}~\bibnamefont {Wang}},
  \bibinfo {author} {\bibfnamefont {H.}~\bibnamefont {Hodovanets}}, \bibinfo
  {author} {\bibfnamefont {H.}~\bibnamefont {Kim}}, \bibinfo {author}
  {\bibfnamefont {D.~J.}\ \bibnamefont {Campbell}}, \bibinfo {author}
  {\bibfnamefont {P.}~\bibnamefont {Zavalij}}, \bibinfo {author} {\bibfnamefont
  {P.}~\bibnamefont {Piccoli}}, \ and\ \bibinfo {author} {\bibfnamefont
  {J.}~\bibnamefont {Paglione}},\ }\href
  {https://doi.org/10.1103%2Fphysrevb.97.224505} {\bibfield  {journal}
  {\bibinfo  {journal} {Phys. Rev. B}\ }\textbf {\bibinfo {volume} {97}}
  (\bibinfo {year} {2018})}\BibitemShut {NoStop}%
\bibitem [{\citenamefont {Lee}\ \emph {et~al.}(2019)\citenamefont {Lee},
  \citenamefont {de~la Pe{\~{n}}a}, \citenamefont {Sun}, \citenamefont
  {Mitrano}, \citenamefont {Fang}, \citenamefont {Jang}, \citenamefont {Lee},
  \citenamefont {Eckberg}, \citenamefont {Campbell}, \citenamefont {Collini},
  \citenamefont {Paglione}, \citenamefont {de~Groot},\ and\ \citenamefont
  {Abbamonte}}]{Lee_2019a}%
  \BibitemOpen
  \bibfield  {author} {\bibinfo {author} {\bibfnamefont {S.}~\bibnamefont
  {Lee}}, \bibinfo {author} {\bibfnamefont {G.}~\bibnamefont {de~la
  Pe{\~{n}}a}}, \bibinfo {author} {\bibfnamefont {S.~X.-L.}\ \bibnamefont
  {Sun}}, \bibinfo {author} {\bibfnamefont {M.}~\bibnamefont {Mitrano}},
  \bibinfo {author} {\bibfnamefont {Y.}~\bibnamefont {Fang}}, \bibinfo {author}
  {\bibfnamefont {H.}~\bibnamefont {Jang}}, \bibinfo {author} {\bibfnamefont
  {J.-S.}\ \bibnamefont {Lee}}, \bibinfo {author} {\bibfnamefont
  {C.}~\bibnamefont {Eckberg}}, \bibinfo {author} {\bibfnamefont
  {D.}~\bibnamefont {Campbell}}, \bibinfo {author} {\bibfnamefont
  {J.}~\bibnamefont {Collini}}, \bibinfo {author} {\bibfnamefont
  {J.}~\bibnamefont {Paglione}}, \bibinfo {author} {\bibfnamefont
  {F.}~\bibnamefont {de~Groot}}, \ and\ \bibinfo {author} {\bibfnamefont
  {P.}~\bibnamefont {Abbamonte}},\ }\href
  {https://doi.org/10.1103%2Fphysrevlett.122.147601} {\bibfield  {journal}
  {\bibinfo  {journal} {Phys. Rev. Lett.}\ }\textbf {\bibinfo {volume} {122}}
  (\bibinfo {year} {2019})}\BibitemShut {NoStop}%
\bibitem [{\citenamefont {Lee}\ \emph {et~al.}(2021)\citenamefont {Lee},
  \citenamefont {Collini}, \citenamefont {Sun}, \citenamefont {Mitrano},
  \citenamefont {Guo}, \citenamefont {Eckberg}, \citenamefont {Paglione},
  \citenamefont {Fradkin},\ and\ \citenamefont {Abbamonte}}]{Lee_2021}%
  \BibitemOpen
  \bibfield  {author} {\bibinfo {author} {\bibfnamefont {S.}~\bibnamefont
  {Lee}}, \bibinfo {author} {\bibfnamefont {J.}~\bibnamefont {Collini}},
  \bibinfo {author} {\bibfnamefont {S.~X.~L.}\ \bibnamefont {Sun}}, \bibinfo
  {author} {\bibfnamefont {M.}~\bibnamefont {Mitrano}}, \bibinfo {author}
  {\bibfnamefont {X.}~\bibnamefont {Guo}}, \bibinfo {author} {\bibfnamefont
  {C.}~\bibnamefont {Eckberg}}, \bibinfo {author} {\bibfnamefont
  {J.}~\bibnamefont {Paglione}}, \bibinfo {author} {\bibfnamefont
  {E.}~\bibnamefont {Fradkin}}, \ and\ \bibinfo {author} {\bibfnamefont
  {P.}~\bibnamefont {Abbamonte}},\ }\href
  {https://doi.org/10.1103/PhysRevLett.127.027602} {\bibfield  {journal}
  {\bibinfo  {journal} {Phys. Rev. Lett.}\ }\textbf {\bibinfo {volume} {127}},\
  \bibinfo {pages} {027602} (\bibinfo {year} {2021})}\BibitemShut {NoStop}%
\bibitem [{\citenamefont {Perdew}\ \emph {et~al.}(1996)\citenamefont {Perdew},
  \citenamefont {Burke},\ and\ \citenamefont {Ernzerhof}}]{PBE}%
  \BibitemOpen
  \bibfield  {author} {\bibinfo {author} {\bibfnamefont {J.~P.}\ \bibnamefont
  {Perdew}}, \bibinfo {author} {\bibfnamefont {K.}~\bibnamefont {Burke}}, \
  and\ \bibinfo {author} {\bibfnamefont {M.}~\bibnamefont {Ernzerhof}},\ }\href
  {\doibase 10.1103/physrevlett.77.3865} {\bibfield  {journal} {\bibinfo
  {journal} {Phys. Rev. Lett.}\ }\textbf {\bibinfo {volume} {77}},\ \bibinfo
  {pages} {3865} (\bibinfo {year} {1996})}\BibitemShut {NoStop}%
\bibitem [{\citenamefont {Kresse}\ and\ \citenamefont {Joubert}(1999)}]{VASP3}%
  \BibitemOpen
  \bibfield  {author} {\bibinfo {author} {\bibfnamefont {G.}~\bibnamefont
  {Kresse}}\ and\ \bibinfo {author} {\bibfnamefont {D.}~\bibnamefont
  {Joubert}},\ }\href {\doibase 10.1103/physrevb.59.1758} {\bibfield  {journal}
  {\bibinfo  {journal} {Phys. Rev. B}\ }\textbf {\bibinfo {volume} {59}},\
  \bibinfo {pages} {1758} (\bibinfo {year} {1999})}\BibitemShut {NoStop}%
\bibitem [{\citenamefont {Bl{\"o}chl}(1994)}]{VASP1}%
  \BibitemOpen
  \bibfield  {author} {\bibinfo {author} {\bibfnamefont {P.~E.}\ \bibnamefont
  {Bl{\"o}chl}},\ }\href {\doibase 10.1103/physrevb.50.17953} {\bibfield
  {journal} {\bibinfo  {journal} {Phys. Rev. B}\ }\textbf {\bibinfo {volume}
  {50}},\ \bibinfo {pages} {17953} (\bibinfo {year} {1994})}\BibitemShut
  {NoStop}%
\bibitem [{\citenamefont {Kresse}\ and\ \citenamefont {Hafner}(1993)}]{VASP2}%
  \BibitemOpen
  \bibfield  {author} {\bibinfo {author} {\bibfnamefont {G.}~\bibnamefont
  {Kresse}}\ and\ \bibinfo {author} {\bibfnamefont {J.}~\bibnamefont
  {Hafner}},\ }\href {\doibase 10.1103/physrevb.47.558} {\bibfield  {journal}
  {\bibinfo  {journal} {Phys. Rev. B}\ }\textbf {\bibinfo {volume} {47}},\
  \bibinfo {pages} {558} (\bibinfo {year} {1993})}\BibitemShut {NoStop}%
\bibitem [{\citenamefont {Kresse}\ and\ \citenamefont
  {Furthm{\"u}ller}(1996)}]{VASP4}%
  \BibitemOpen
  \bibfield  {author} {\bibinfo {author} {\bibfnamefont {G.}~\bibnamefont
  {Kresse}}\ and\ \bibinfo {author} {\bibfnamefont {J.}~\bibnamefont
  {Furthm{\"u}ller}},\ }\href {\doibase 10.1103/physrevb.54.11169} {\bibfield
  {journal} {\bibinfo  {journal} {Phys. Rev. B}\ }\textbf {\bibinfo {volume}
  {54}},\ \bibinfo {pages} {11169} (\bibinfo {year} {1996})}\BibitemShut
  {NoStop}%
\bibitem [{\citenamefont {Becke}\ and\ \citenamefont {Edgecombe}(1990)}]{elf}%
  \BibitemOpen
  \bibfield  {author} {\bibinfo {author} {\bibfnamefont {A.~D.}\ \bibnamefont
  {Becke}}\ and\ \bibinfo {author} {\bibfnamefont {K.~E.}\ \bibnamefont
  {Edgecombe}},\ }\href {https://doi.org/10.1063/1.458517} {\bibfield
  {journal} {\bibinfo  {journal} {J. Chem. Phys.}\ }\textbf {\bibinfo {volume}
  {92}},\ \bibinfo {pages} {5397} (\bibinfo {year} {1990})}\BibitemShut
  {NoStop}%
\bibitem [{\citenamefont {Blaha}\ \emph {et~al.}(2001)\citenamefont {Blaha},
  \citenamefont {Schwarz}, \citenamefont {Madsen}, \citenamefont {Kvasnicka},\
  and\ \citenamefont {Luitz}}]{WIEN2k}%
  \BibitemOpen
  \bibfield  {author} {\bibinfo {author} {\bibfnamefont {P.}~\bibnamefont
  {Blaha}}, \bibinfo {author} {\bibfnamefont {K.}~\bibnamefont {Schwarz}},
  \bibinfo {author} {\bibfnamefont {G.~K.~H.}\ \bibnamefont {Madsen}}, \bibinfo
  {author} {\bibfnamefont {D.}~\bibnamefont {Kvasnicka}}, \ and\ \bibinfo
  {author} {\bibfnamefont {J.}~\bibnamefont {Luitz}},\ }\href@noop {} {\emph
  {\bibinfo {title} {WIEN2k, An Augmented Plane Wave+Local Orbitals Program for
  Calculating Crystal Properties}}}\ (\bibinfo  {publisher} {K. Schwarz, Tech.
  Univ. Wien, Austria},\ \bibinfo {year} {2001})\BibitemShut {NoStop}%
\bibitem [{\citenamefont {Madsen}\ and\ \citenamefont
  {Singh}(2006)}]{Madsen_2006}%
  \BibitemOpen
  \bibfield  {author} {\bibinfo {author} {\bibfnamefont {G.~K.~H.}\
  \bibnamefont {Madsen}}\ and\ \bibinfo {author} {\bibfnamefont {D.~J.}\
  \bibnamefont {Singh}},\ }\href {\doibase 10.1016/j.cpc.2006.03.007}
  {\bibfield  {journal} {\bibinfo  {journal} {Comput. Phys. Commun.}\ }\textbf
  {\bibinfo {volume} {175}},\ \bibinfo {pages} {67} (\bibinfo {year}
  {2006})}\BibitemShut {NoStop}%
\bibitem [{\citenamefont {Hoffmann}\ and\ \citenamefont
  {Zheng}(1985)}]{hoffmann}%
  \BibitemOpen
  \bibfield  {author} {\bibinfo {author} {\bibfnamefont {R.}~\bibnamefont
  {Hoffmann}}\ and\ \bibinfo {author} {\bibfnamefont {C.}~\bibnamefont
  {Zheng}},\ }\href {https://doi.org/10.1021/j100266a007} {\bibfield  {journal}
  {\bibinfo  {journal} {J. Phys. Chem.}\ }\textbf {\bibinfo {volume} {89}},\
  \bibinfo {pages} {4175} (\bibinfo {year} {1985})}\BibitemShut {NoStop}%
\bibitem [{\citenamefont {Cao}\ \emph {et~al.}(2013)\citenamefont {Cao},
  \citenamefont {Chakoumakos}, \citenamefont {Chen}, \citenamefont {Yan},
  \citenamefont {{McGuire}}, \citenamefont {Yang}, \citenamefont {Custelcean},
  \citenamefont {Zhou}, \citenamefont {Singh},\ and\ \citenamefont
  {Mandrus}}]{cao-irte2}%
  \BibitemOpen
  \bibfield  {author} {\bibinfo {author} {\bibfnamefont {H.}~\bibnamefont
  {Cao}}, \bibinfo {author} {\bibfnamefont {B.~C.}\ \bibnamefont
  {Chakoumakos}}, \bibinfo {author} {\bibfnamefont {X.}~\bibnamefont {Chen}},
  \bibinfo {author} {\bibfnamefont {J.}~\bibnamefont {Yan}}, \bibinfo {author}
  {\bibfnamefont {M.~A.}\ \bibnamefont {{McGuire}}}, \bibinfo {author}
  {\bibfnamefont {H.}~\bibnamefont {Yang}}, \bibinfo {author} {\bibfnamefont
  {R.}~\bibnamefont {Custelcean}}, \bibinfo {author} {\bibfnamefont
  {H.}~\bibnamefont {Zhou}}, \bibinfo {author} {\bibfnamefont {D.~J.}\
  \bibnamefont {Singh}}, \ and\ \bibinfo {author} {\bibfnamefont
  {D.}~\bibnamefont {Mandrus}},\ }\href
  {https://doi.org/10.1103/PhysRevB.88.115122} {\bibfield  {journal} {\bibinfo
  {journal} {Phys. Rev. B}\ }\textbf {\bibinfo {volume} {88}},\ \bibinfo
  {pages} {115122} (\bibinfo {year} {2013})}\BibitemShut {NoStop}%
\bibitem [{\citenamefont {Li}\ \emph {et~al.}(2014)\citenamefont {Li},
  \citenamefont {Lin}, \citenamefont {Yan}, \citenamefont {Chen}, \citenamefont
  {Gianfrancesco}, \citenamefont {Singh}, \citenamefont {Mandrus},
  \citenamefont {Kalinin},\ and\ \citenamefont {Pan}}]{li-irte2}%
  \BibitemOpen
  \bibfield  {author} {\bibinfo {author} {\bibfnamefont {Q.}~\bibnamefont
  {Li}}, \bibinfo {author} {\bibfnamefont {W.}~\bibnamefont {Lin}}, \bibinfo
  {author} {\bibfnamefont {J.}~\bibnamefont {Yan}}, \bibinfo {author}
  {\bibfnamefont {X.}~\bibnamefont {Chen}}, \bibinfo {author} {\bibfnamefont
  {A.~G.}\ \bibnamefont {Gianfrancesco}}, \bibinfo {author} {\bibfnamefont
  {D.~J.}\ \bibnamefont {Singh}}, \bibinfo {author} {\bibfnamefont
  {D.}~\bibnamefont {Mandrus}}, \bibinfo {author} {\bibfnamefont {S.~V.}\
  \bibnamefont {Kalinin}}, \ and\ \bibinfo {author} {\bibfnamefont
  {M.}~\bibnamefont {Pan}},\ }\href {https://doi.org/10.1038/ncomms6358}
  {\bibfield  {journal} {\bibinfo  {journal} {Nat. Commun.}\ }\textbf {\bibinfo
  {volume} {5}},\ \bibinfo {pages} {5358} (\bibinfo {year} {2014})}\BibitemShut
  {NoStop}%
\end{thebibliography}%

\end{document}